\documentclass[journal,doublecolumn,9pt]{IEEEtran}
\usepackage{setspace}
\usepackage{epsfig,latexsym}
\usepackage{float}
\usepackage{indentfirst}
\usepackage{amsmath, amssymb}
\usepackage{times}
\usepackage{subfigure}
\usepackage{psfrag}
\usepackage{cite}

\usepackage{multicol}
\usepackage{graphicx}

\usepackage{commands}
\usepackage{matrix}
\usepackage{vect}

\usepackage[dvips]{color}

\usepackage{algorithm}
\usepackage{algpseudocode}

\newcommand{\tra}{{\mathrm{\scriptscriptstyle T}}}

\newcommand{\mE}{\mbox{E}}
\newcommand{\md}{\mbox{d}}

\begin{document}

\title{Hybrid Demodulate-Forward Relay Protocol for Two-Way Relay Channels}

\author{ \authorblockN{Chunbo Luo$^\dagger$, Cathryn Peoples$^\ddagger$, Gerard Parr$^\ddagger$, Sally McClean$^\ddagger$ and  Xinheng Wang$^\dagger$\\}
\authorblockA{ $\dagger$ School of Engineering and Computing, University of the West of Scotland, Paisley, PA1 2BE, UK.\\
$\ddagger$School of Computing and Information Engineering, 
University of Ulster, Coleraine, BT52 1SA, UK.\\
E-mail: {\tt \{chunbo.luo, xinheng.wang\}@uws.ac.uk, \{c.peoples, gp.parr, si.mcclean\}@ulster.ac.uk} \\
\vspace*{0mm}}}

\begin{singlespace}

\maketitle

\thispagestyle{empty}

\end{singlespace}

%\vspace{2em}

\begin{abstract}

Two-Way Relay Channel (TWRC) plays an important role in relay networks, and efficient relaying protocols are particularly important for this model. {\color{black} However, existing protocols may not be able to realize the potential of TWRC if the two independent fading channels are not carefully handled}. In this paper, a Hybrid DeModulate-Forward (HDMF) protocol is proposed to address such a problem. {\color{black} We first introduce the two basic components of HDMF - direct and differential DMF, and then propose the key decision criterion for HDMF based on the corresponding log-likelihood ratios. We further enhance the protocol so that it can be applied independently from the modulation schemes. Through extensive mathematical analysis, theoretical performance of the proposed protocol is investigated. By comparing with existing protocols, the proposed HDMF has lower error rate. A novel scheduling scheme for the proposed protocol is introduced, which has lower length than the benchmark method. The results also reveal the protocol's potential to improve spectrum efficiency of relay channels with unbalanced bilateral traffic.}
\end{abstract}


%=== Section Introduction ===%

\section{Introduction}\label{secInt}

% Relay channel and relaying procotols %

\begin{figure}[tb]\centering
\centering
\epsfig{file=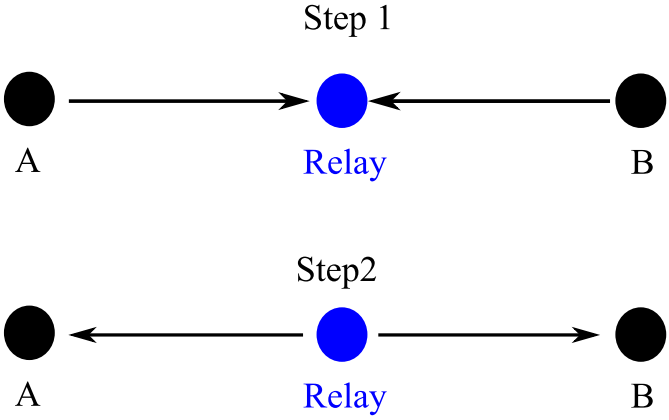,width=1\columnwidth}
\caption{Two-way relay channel model.}
\label{figNetworkStructure}
\end{figure}

A typical Two-Way Relay Channel (TWRC) involves two source nodes (A and B) exchanging data through one intermediate relay (Relay) \cite{Zhiyong2014SEC,Chen2014FTW,Deze2014PDS}, as shown in Fig.\ref{figNetworkStructure}. The two-way problem was first studied by Shannon \cite{Shannon61} and analyzed theoretically in terms of capacity by \cite{Cover79}. Recent research confirms that the employment of relays in certain conditions can significantly increase the performance of wireless networks if they are handled effectively using suitable relaying protocols \cite{Laneman04,Sendonaris03a}. To this end, several common relaying protocols, e.g., Amplify-Forward (AF) and Decode-Forward (DF) were proposed. However, AF has the problem of noise amplification and involves expensive RF chains to mitigate the existing coupling effects, thus practical DF protocols have been proposed which, through proper demodulator design, can also have better performance and achieve full diversity \cite{Wang07HPC}.

% and DF has higher complexity and can propagate decoding errors. In order to overcome the drawbacks of these protocols, DeModulate-Forward (DMF) protocol \cite{Chen06} has been proposed by the authors which reduces noise by demodulation and has lower complexity compared with DF. 

% Existing solutions / why do they not work?

The TWRC model has extensive real-world instances, such as ground stations connected by satellites, two mobile users exchanging information through one common base station or relay, or sensors connected by a smart router, for example \cite{Falaki10Traff,Dong00Asym,Tanenbaum2003computer}. In our recent project on developing UAVs to bridge communications between wireless users, such a model can be directly adopted \cite{suaave10, ChunboACM2014}. As the two source nodes can transmit signals simultaneously, an effective relaying protocol with high efficiency is critical to system capacity. However, the application of some simple protocols, e.g. AF/DF, incurs low bandwidth efficiency because four time slots are required to finish the exchange of one message from each source. 

% DNC/ANC

Considering the special structure of TWRC, researchers have applied the concept of network coding in the wireless domain and proposed several protocols that need only two or three time slots. As a result, significant spectrum usage (or time)-at least $50\%$-can be saved. Such protocols include Analog Network Coding (ANC) \cite{Katti07embracingwireless}, Physical-layer Network Coding (PNC) \cite{Shengli07}, Digital Network Coding (DNC) \cite{Chen13DNC} and their equivalences. Even though these new protocols can theoretically improve system throughput by up to $100\%$ over the direct relaying transmission scheme, due to the limit of channel effects such as fading, Signal-to-Noise Ratio (SNR) variation and system implementation difficulties such as imperfect synchronization, the claimed advantages cannot be fully achieved in real-life scenarios. For example, if channel gains of the two source-relay channels have large discrepancies, these schemes may incur large packet error rates and spectrum efficiency is greatly decreased by retransmitting large amounts of erroneous packets \cite{Gong12HDMF, Shengli07}. This problem is particularly severe for the PNC protocol as it achieves the best performance when both channel gains are the same. It is therefore of theoretical interest and practical importance to design protocols/network coding schemes which can overcome the problems of wireless channels with dynamic fading and large SNR variations. { \color{black} This is the problem which this paper wishes to solve. }

{\color{black}
Current efforts on the differential demodulation of received superposed signals to reduce noises have been substantially studied in the literatures\cite{Noori2012OSM,Anxin2008AED,Koike2009OCF, Hyun2010MHO}. Such modulation is an example of the compute-forward protocol used for PNC.} The authors of \cite{Noori2012OSM} summarized the necessary and sufficient conditions on the mapping of symbols at the user side to remove the ambiguity at relay, and thus improved the error rate performance of a system adopting phase-shift keying modulations. { \color{black} A novel DeNoise-and-Forward (DNF) scheme was proposed in \cite{Anxin2008AED}, which addressed the noise accumulation issue at the relay node of a two-way system and inspired the proposed modulation-independence protocol of this paper.}  The analysis from \cite{Koike2009OCF} shows that at the source to relay phase, the traditional operation of PNC - ExclusiveOR  does not always offer the best performance, thus the authors proposed a design strategy to optimize constellations for improved throughput. {\color{black} Similar focus on the constellation design can also be found in \cite{Hyun2010MHO} which aimed at removing ambiguity points of M-PAM (pulse amplitude modulation) signal constellations to enable using binary physical-layer network coding in an efficient way. This paper also provided an instrument for higher order modulation schemes.} Even though extensive studies have been carried out, practical scenarios with variable channels and large SNR variations with the focus on low error rate and queue scheduling are still yet to be explored and leave the gap for further work.    

% Background / Significance
%Particularly, it is often observed in networks that the bidirectional channel gains are not all the same\cite{Falaki10Traff,Dong00Asym}, the intermediate relay node sees signals with different strengths. On one hand, network coding could be used to improve the system throughput, however, it has high error rate if the channel gains of the two routes (e.g. A-R and B-R) have large discrepancy and the Signal-Noise Ratio (SNR) and would incur frequent retransmission thus decrease the channel efficiency. 

%Even in traditional wireless/wired networks, small volumes of backward feedback messages should be sent back to the transmitter for the purpose of performance monitoring and automatic repeat query - this can occur even in the absence of application traffic on the forward pass. 

% significant advancements have been achieved in the areas of relay networks \cite{Laneman04}, cooperative communications \cite{Sendonaris03a}, network coding \cite{Li03}. These contributions greatly enhance the quality of wireless communications and provide opportunities for mobile devices to be connected at any time and anywhere. In particular, relays show advantages in providing higher quality communication and extending network coverage across data networks \cite{Laneman04}. 

% Other related work

%1. Their focuses and proposals: 2. Their drawbacks: only one modulation types; waster of energy on processing; increased complexity;

% Details of the scheme

In this paper, we propose a Hybrid DeModulate-Foward (HDMF) protocol specialized for the TWRC model. Specifically, the HDMF protocol focuses on two-way relay channels which do not always have equal channel gains because of fading and synchronization errors. The aim of this protocol is to achieve both high spectrum efficiency and low packet error rate when compared with the typical protocols introduced previously. HDMF incorporates differential and direct DMF, and adapts its strategy based on the channel coefficients to reduce packet errors. A key to this protocol is the decision rule which regulates the two DMF schemes. The corresponding Log-Likelihood Ratios (LLR) are used to form the decision criterion. We also improve the simple HDMF protocol for the case where high dimensional modulation schemes are used. The proposed HDMF does not only increase spectrum efficiency over direct relay schemes, but also overcome the limits of ANC/PNC schemes. {\color{black} The proposed HDMF can be directly extended to the hybrid-DF case, however, since this only increases the complexity (by adding a decoder) without bringing more understanding about the system, hybrid-DF is not covered in this paper.} 

% Further details

Furthermore, we propose a scheduling scheme to adjust the data flows of HDMF considering dynamic channel conditions. We introduce four transmission modes and develop a scheduling scheme based on the transition probabilities of the modes. Without loss of generality, we only study the transition states of the queue length at source node A and the forwarding queue length from A to B at Relay. Similar behaviour can be observed at B and also the queue at Relay (from B to A), thus they are omitted in the paper. Queue behaviour is summarized for all four modes and is an essential component of the scheduling scheme: The average length of the queue at A and the forward queue at Relay for data to B are investigated based on a Markov chain model. Its performance is verified using simulation and compared with the DNC scheme studied in \cite{Chen13DNC}. 

% contributions

The main contributions of this paper are summarized as follows. 
\begin{itemize}
\item A HDMF protocol proposal for the TWRC model with fading channels and variable SNR; 
\item A scheduling scheme proposal for the TWRC model with the HDMF protocol;
\item A study of average queue lengths of the scheduling scheme with the HDMF protocol. 
\end{itemize}

% Paper structure

This paper continues as follows: Section \ref{secMod} describes the system model; Section \ref{secHDMF} proposes the hybrid DMF protocol; Section \ref{secTSF} designs the scheduling scheme and investigates the queue length transition states; Section \ref{secSim} verifies its performance by simulation;  Section \ref{secDis} discusses the important issues; Section \ref{secCon} concludes the paper.

\section{The System Model}\label{secMod}

The system model (Fig. \ref{figNetworkStructure}) considered in this work is the typical TWRC model with two source nodes to exchange information through an intermediate relay. The channels are assumed to remain unchanged for at least one packet duration and the coefficients are known by the corresponding destinations. In designing the protocol, we assume noise to have a Gaussian distribution with zero mean and $N_0/2$ variance. 

%There may exist a direct link between S and D and it may have positive effect on the system performance; in order to concentrate on relay protocols, the main issue in this paper, the S-D link is ignored here. 

At every time slot, sources A and B simultaneously transmit data packets, e.g., $\x_a(n)$ and  $\x_b(n)$ to Relay. Each packet has $M$ symbols, denoted as $\x_a(n) = [x_a^1(n),\cdots,x_a^M(n)]^\tra$, $\x_b(n) = [x_b^1(n),\cdots,x_b^M(n)]^\tra$. Relay receives superposed versions of the two packets from A and B as follows
\begin{equation}\label{ydyr}
\y_{r}(n)=h_{AR}(n) \x_a(n) + h_{BR}(n) \x_b(n)+\w_{r}(n),
\end{equation}
where $\y_r(n) = [y_r^1(n),\cdots,y_r^M(n)]^\tra$, $h_{AR}(n)$ and $h_{BR}(n)$ are the channel coefficients between A and R, B and R at the time slot $n$. $\w_r(n)$ is the noise vector at Relay. The $m$th symbol in $\y_r(n)$ can be denoted as $y_r^m(n)$.

Messages received at Relay are processed and forwarded to the corresponding destinations by protocols, such as DMF, AF and DF. The forwarded signal $\x_r(n)$ is given by
\begin{equation}\label{xr}
\x_{r}(n)=f(\y_r(n-1) ),
\end{equation}
where $\y_r(n-1)$ is the received signal by Relay at the $(n-1)$th time slot and $f( \cdot )$ is the relaying function which describes the processes of the relaying protocol. {\color{black}  For example, if using direct DMF \cite{Gong12HDMF}, $f(\cdot)$ denotes the process of demodulating the received signal and then re-modulating it using the designated modulation scheme. In this case, data from the other user will not be detected. { \color{black} However, such case only happens if the channel quality deteriorates to a considerably low level as denoted by the likelihood ratio value.  In order to ensure the dropped data can be retransmitted, current communication systems usually adopt Automatic Repeat reQuest (ARQ) schemes which are also assumed in this paper. The implementation details of ARQ are omitted here and can be found in \cite{Shu1984ARQ}.} If the weaker channel becomes strong enough, differential DMF will be automatically switched, where $f(\cdot)$ denotes the demodulation and remodulation of the data from both users. Even though sequential interference cancellation techniques can be used, it is not the focus here since the error rate of the weaker channel would be even higher than the stronger one.} 

The received signals at source B and A at the $n$th time slot are given
by
\begin{equation}\label{yd_Dl}
\begin{array}{l}
\y_b(n)=h_{RB}(n) \x_{r}(n)+\w_b(n), \\
\y_a(n)=h_{RA}(n) \x_{r}(n)+\w_a(n), \\
\end{array}
\end{equation}
where $\y_b(n) = [y_b^1(n),\cdots,y_b^M(n)]^\tra$, $h_{RB}(n)$ is the channel coefficient between $R$ and $B$ at the $n$th time slot, and $\w_b(n)$ is the noise vector at node B. $\y_a(n)$ and $h_{RA}(n)$ have similar definitions as those for source B. 

%Similarly, S receives the destination feedback message from R as
%\begin{equation}\label{yS}
%\begin{array}{l}
%\y_S(n)=h_{RA}(n) \x_{R}(n)+\w_S(n).
%\end{array}
%\end{equation}
%This feedback message is then used for further processing. It is worth mentioning that the link for feedback messages allow lower data rates and greater tolerance towards delay and connection quality in this model. Throughout this paper, the name of forward and backward route is also used to denote the main data and feedback route respectively. 

% === HDMF === %

\section{Hybrid Demodulation-Forward Protocol}\label{secHDMF}

{\color{black}

This section proposes the HDMF protocol. The general idea of HDMF is to have two DMF modes including direct DMF and differential DMF, and switch the two modes automatically based on a criterion obtained from the channel gains and SNRs. The protocol is implemented on the Relay and destinations respectively, where the Relay uses it to construct and forward signals. After the reception of these signals, A and B apply the protocol to detect the source data. We use Quadrature Phase Shift Keying (QPSK) as an example to explain the mechanism of this protocol which can be easily extended to higher-order modulation scenarios. The details are elaborated in the final part of this section. 
}  
%We further analyse the ratio of packets sending through differential and direct DMF and have a deep understanding of the HDMF protocol. 

%The focus of HDMF is on how to process signals received by relay and the detection at end notes. 
%
%Classic DMF \cite{Chen06} relay system directly demodulates the source data, remodulates and forwards it to the destination. As shown in \eqref{ydyr}, the performance of the direct demodulation can be severely limited if the source and feedback signal have comparable power. We can differentially demodulate the two signals and making good use of the comparable power, thus providing better performance. 

To introduce the principles of this protocol: we assume that data packets from the two source nodes are of the same length. (The case of unequal length packets will be discussed in Section \ref{secDis}.) Each packet has $M$ symbols, and each symbol denotes $K$ bits if the modulation scheme has $K$-th order. For example, the source node packet $\x_a(n)$ has $M$ symbols and the $m$th symbol $x^m_a(n)$ has $K$ bits which are denoted as $\{b^m_{a,1}(n),\cdots,b^m_{a,K}(n)\}$. Similarly, $x^m_b(n)$ has $K$ bits and are denoted as $\{b^m_{b,1}(n),\cdots,b^m_{b,K}(n)\}$. We assume the symbols are generated randomly with equal probabilities. Therefore, for Quadrature Phase Shift Key (QPSK) modulation, each symbol has two bits since $K=2$, and the four different constellation symbols of QPSK are generated randomly with the same probability: $\frac{1}{4}$.  

{\color{black}
The following subsections first introduce the two component parts of the HDMF protocol, and then present the decision criterion to choose these two components and the implementation details.
}

%Thus the bit index $k$ can be ignored without causing confusion since there is only one bit per symbol. All the symbols are randomly generated with equal probabilities, in the case of BPSK, $P(b_S(n_m) = 1)=P(b_S(n_m) = 0)=1/2$. 

\subsection{Direct Demodulation-Forward}

The received signal at Relay is shown in \eqref{ydyr}. As two signals arrive at Relay simultaneously, we should try to demodulate the stronger signal directly in order to control the detection errors if one channel has significantly higher channel gain than the other and the noise is strong. Without loss of generality, we take the route from A to Relay to B as an example. 

\begin{figure}[tb]\centering
    \epsfig{file=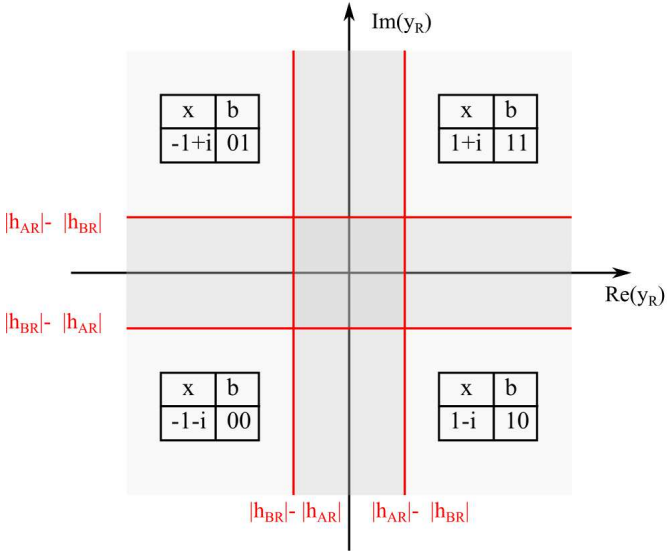, width=1\columnwidth,clip=}
    \caption{ {\color{black} The demodulation scheme of direct DMF where $|h_{AR}|\geq| h_{BR}|$} and QPSK. The dark shadow areas around the two decision lines denote high detection error probabilities. The corresponding symbols and data bits are mapped using Gray code and  shown in the four corners.} 
    \label{figDirectDMF}
\end{figure}

The maximum likelihood (ML) criterion is used to demodulate the $m$th symbol within $\x_a(n)$, 
\begin{equation} \label{eq:xsml}
\hat x^m_a(n)  = {\rm arg} \max_{x^m_a(n)\in \mathcal{M}} \left\{ P(y^m_r(n)|x^m_a(n)  \right\}, \; m=1,...,M.
\end{equation}
where $\mathcal{M}$ is the modulation symbol set. For QPSK, $\mathcal{M} = \{1+i, -1+i, -1-i, 1-i \}$. Therefore, direct DMF has the forwarding signal as follows
\begin{equation}
\x_r(n) = \hat \x_a (n) = \left\{ \hat x^1_a(n), ...,\hat x^M_a(n)  \right\} ^\tra
\end{equation}

\emph{In order to describe the scheme concisely, the symbol index $m$ is omitted henceforward and $x_a$ always denotes the $m$th symbol within $\x_a$. }

Since all symbols are drawn under the same probability, from \eqref{ydyr} we have
\begin{equation} \label{eq:bslpyi}
\begin{aligned}
P&(y_r(n)| x_a(n))\\
 &= \sum_{x_b(n) \in \mathcal{M}} P(x_b(n)) \cdot P(y_r(n)| {x}_a(n), x_b(n)). 
 \end{aligned}
\end{equation}

Given QPSK modulation, $P(x_b(n))=1/4, \; x_b(n) \in \mathcal{M}$. \eqref{eq:bslpyi} can be simplified as
\begin{equation} \label{eq:bsl}
\begin{aligned}
P&(y_r(n)| x_a(n))\\
 &= \frac{1}{4} \sum_{x_b(n) \in \mathcal{M}} P(y_r(n)| {x}_a(n), x_b(n)).
\end{aligned}
\end{equation}
Under the ML criterion, the symbol ($x_a(n)$) with the maximum $P(y_r(n)| x_a(n))$ value is chosen to be the detected symbol within $\x_r(n)$.  

%\subsection{When should direct DMF be used?}

\begin{figure}[tb]\centering
    \epsfig{file=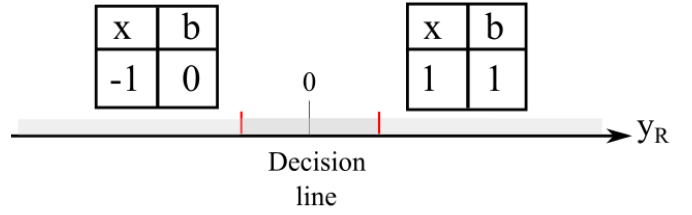, width=\columnwidth}
    \caption{The demodulation scheme of direct DMF under BPSK modulation.} 
    \label{figDmfBpsk}
\end{figure}

{\color{black}

Fig. \ref{figDirectDMF} shows the decision areas of the direct DMF scheme for the case of $|h_{AR}|\geq |h_{BR}|$ under QPSK modulation. The other case when $|h_{AR}| < |h_{BR}|$ can be similarly analyzed and is omitted here. In Fig.\ref{figDirectDMF}, the decision lines are the axes. The detected symbols and the corresponding data bits are given in the tables. The areas are derived based on the distances between the received signal position and the decision lines for illustration purposes. (The decision areas for Binary Phase-Shift Keying (BPSK) modulation is also given here and shown in Fig.\ref{figDmfBpsk}, which is a simpler version of the QPSK case.)
}

Ideally, the received symbols should be located at the four constellation points in the quadrants of the Cartesian system. However, because of noise and interference from the other channel, the received signals of $y_r$ are scattered on the plane around the four lines ($|h_{AR}(n)| \pm  |h_{BR}(n)|$, $ |h_{AR}(n) \pm h_{BR}(n)| \times i $) which form the two stripes in the figure. If some received symbols fall into the dark areas, a small noise can push them to the other side of the decision line, which will then be interpreted as different symbols and cause detection errors. For symbols located in the lighter areas, because they are far away from the decision lines, they need stronger noise (or lower SNR) to cause demodulation errors.  

Since the received signals are around the above four lines, in order to have less detection error, the lines should be far away from the axes so that even though some symbols are located in the dark areas, they would still be far from the decision lines and are strong enough to combat noise. Therefore the width of the dark stripes is essential for direct DMF. From \eqref{ydyr}, we can see that this width is controlled by the relevant channel gains of the two channels - the larger the discrepancies between the two channels, the wider the stripes in the figure. The ideal case to apply direct DMF is when one channel is much stronger than the other so that we have large absolute value of $|h_{AR}(n)| - |h_{BR}(n)| $, thus the lines are far away from the axes.

By way of contrast, we can see where direct DMF will introduce problems by looking at an opposite example where the two channels have equal channel coefficients. The stripe width will be 0 and the received symbols will be around the axes (which are the decision lines) and the  lines with the value of double channel gains. We can therefore expect $\frac{3}{8} $ of error rate because a small noise can push the received signals around the axes  from one side to the other side. The clear boundaries where direct DMF should be applied are essential to the HDMF protocol and should be jointly considered with SNR. Before we discuss this subject, we introduce a solution to the scenarios where the direct DMF fails. 

% === Differential DmF === %

\subsection{Differential Demodulation-Forward } \label{secDifDMF}

Previous discussion reveals that direct DMF would generate results with high error rates if the two channels of TWRC have comparable gains, e.g. $|h_{AR}(n)|\approx |h_{BR}(n)|$. In order to solve this problem, we propose to use differential DMF rather than direct DMF in some scenarios. Fundamentals of this scheme are introduced using QPSK as an example. 

\subsubsection{Differential DMF at Relay}

In order to demonstrate the differential DMF algorithm, we firstly consider an instance where there is no noise and $h_{AR}(n)=h_{BR}(n)$. In this case, the two source signals of \eqref{ydyr} arrive with the same power at Relay. The received superposed signal can be denoted as 
\begin{equation}\label{yrDifDMF}
\y_{r}(n)=h_{AR}(n) (\x_a(n)+\x_b(n)).
\end{equation}
For a given modulation scheme like QPSK, both the real and imaginary part of $y_r(n)$ only have two possible absolute values: $2|h_{AR}(n)|$ and $0$ respectively. These two values correspond to the two input sets: $b_{a,k}(n) = b_{b,k}(n)$ and $b_{a,k}(n) \neq b_{b,k}(n)$. It is easy for the relay to directly demodulate the received symbols at the first instance as both of them are the same and the power is enhanced. However, the latter case adds difficulty as it may be the superposition of two different bit sets: $b_{a,k}(n)= 1,\; b_{b,k}(n)=0$ and $b_{a,k}(n)= 0,\; b_{b,k}(n)=1$, ($k=1,2$). However, since both the two sources know their transmitted signals, it is desirable for both A and B that Relay differentially demodulates and forwards signals to them. 

The differential DMF at Relay, after converting data symbols to data bits, is given as
\begin{equation} \label{eq:bsop}
\hat b_{r,k}(n) = \hat b_{a\oplus b, k}(n) = b_{a,k}(n) \oplus b_{b,k}(n), \; k=1,2,
\end{equation}
where $\oplus$ is the XOR (exclusive OR) operation. For example, if node A has a symbol 1+i and node B has -1+i, the corresponding bits of A's symbol are 1,1 and those of B's symbol are 0,1. The resulting two bits after the above processing can be given as $\hat b_{r,1}(n) = 1 \oplus 0 = 1$ and $\hat b_{r,2}(n) = 1 \oplus 1 = 0$.

The above signal processing can be applied directly in the symbol domain if we can demodulate the differential between $b_{a,k}(n)$ and $b_{b,k}(n)$ as follows
\begin{equation}\label{eqBSxorD}
\begin{aligned}
&\hat b_{a \oplus b,k}(n) = \\
&\left \{
\begin{array}{ll}
0,  {\rm Re} \{y_r(n)\} = 2{\rm Re} \{ h_{AR}(n) \}; & 1, {\rm Re} \{ y_r(n) \} = 0 \\
0,  {\rm Im} \{y_r(n)\} = 2{\rm Im} \{ h_{AR}(n) \}; & 1, {\rm Im} \{ y_r(n)\} = 0
\end{array} \right.
\end{aligned}
\end{equation}
where ${\rm Re\{ \cdot \}}$ and ${\rm Im\{ \cdot \}}$ denote the real and imaginary part of the input, respectively. The first line of \eqref{eqBSxorD} is for the first bit and the second line is for the second bit because one QPSK symbol denotes two bits.

The ML differential detection at relay is given by
\begin{equation} \label{eq:xdiml}
\hat x_{a \oplus b}(n)  = {\rm arg} \max_{x_{a \oplus b}(n)  \in \mathcal{M} } \left\{ P(y_r(n)|x_{a\oplus b}(n) \right\},
\end{equation}
where $P(y_r(n)|x_{a \oplus b}(n) \in \mathcal{M})$ is the distributions of $y_r(n)$ given the differential symbols within modulation set $\mathcal{M}$.  

From \eqref{ydyr}, it can be seen that
\begin{equation} \label{eq:dibsl}
\begin{aligned}
P&(y_r(n)  |x_{a\oplus b}(n)) = \sum_{x_a(n) \in \mathcal{M}, x_b(n) \in \mathcal{M}} \\
& P(x_a(n),x_b(n))  P(y_r(n)|x_a(n) \oplus x_b(n)).
\end{aligned}
\end{equation}
Under the conditions of QPSK modulation and symbols generated with equal probability, 
\begin{equation} \label{eq:dibs0}
\begin{aligned}
P&(y_r(n)  |x_{a\oplus b}(n)) \\
& = \frac{1}{4} \sum_{x_a(n) \in \mathcal{M}, x_b(n) \in \mathcal{M}}  P(y_r(n)|x_a(n) \oplus x_b(n)).
\end{aligned}
\end{equation}

\begin{figure}[tb]\centering
\epsfig{file=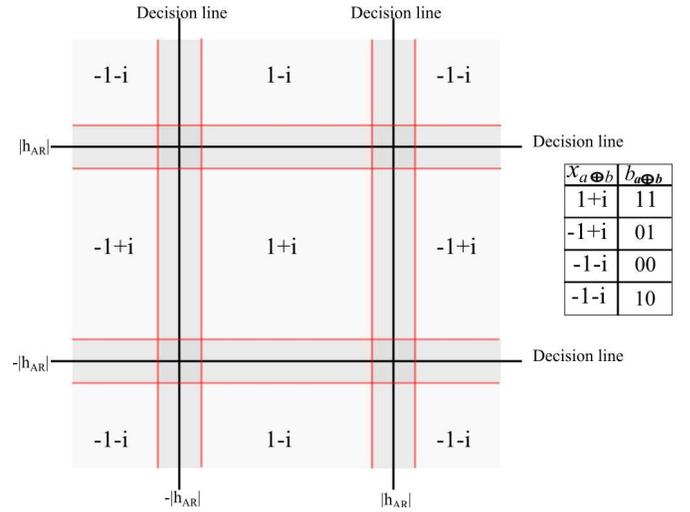, width=1\columnwidth,clip=}
\caption{The demodulation scheme of differential DMF under QPSK modulation. If the received symbols are located at the dark shadow areas, they have potentially high detection errors. { \color{black} (Assume $ |h_{AR}| \geq |h_{BR}|$)}.} 
\label{figDifferentialDMF}
\end{figure}

Relay can demodulate the received signals differentially using the above ML detection algorithm and forward the results to A and B. Take the case of $ |h_{AR}| \geq |h_{BR}|$ for example. The demodulation scheme of differential DMF is shown in Fig. \ref{figDifferentialDMF}. Similar to Fig. \ref{figDirectDMF}, the dark stripes are the areas which has more detection errors than the rest.  The decision lines are formed by $\pm |h_{AR}|$ and $\pm |h_{AR}| i$. The edges are formed by $\pm |h_{AR}| \pm |h_{BR}|$ and their imaginary equivalents. 

\begin{figure}[tb]\centering
    \epsfig{file=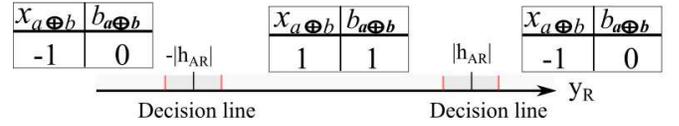, width=\columnwidth}
    \caption{The demodulation scheme of differential DMF under BPSK modulation.} 
    \label{figDifDmfBpsk}
\end{figure}

Similarly, the received symbols are located around the lines which form the four stripes. However, in contrast to direct DMF, the decision lines are not the two axes anymore. Instead, they are $\pm |h_{AR}|$ and $\pm |h_{AR}| i$. In this case, the symbols which are located around the axes can mostly combat higher noise. The ideal case is when the two channel gains are the same, e.g. $|h_{AR}| = |h_{BR}|$, (where direct DMF has the worst performance) so that the majority of the received symbols will be located around the axes (not depicted in the figure) and far away from the decision lines. In this case, they will strongly combat noise. The width of the stripes will decrease if the two channels have different gains.  The worst case is when one of them is zero, e.g. $|h_{BR}| = 0$ in Fig. \ref{figDifferentialDMF}. The width, in this case, will become zero and the edges (mean of the received symbols) overlap with the decision lines. In this case, a small noise will push the signals from the correct side to the wrong side, thus generating a large error rate of $\frac{3}{8}$. Another important job is to decide when to use differential DMF. (Fig.\ref{figDifDmfBpsk} gives the decision areas for the similar case under BPSK modulation, from which we can see when $y_R$ has values around the two decision lines: $-|h_{AR}|$ and $|h_{AR}|$, the detected symbol has a higher error probability.)

\subsubsection{Detection of differential DMF symbols at A and B}\label{secDET}

The signal received by the two terminal nodes A and B are given by \eqref{yd_Dl}. We use node A, for example: $\y_a(n) = h_{RA}(n) \x_r(n)+\w_a(n)$ and $\x_r (n) = f(y_r(n-1))$ (given by \eqref{xr}). At the $n$th time slot, if the differential DMF protocol is applied to generate $\x_r(n)$, we obtain $\hat b_{r,k}(n) = \hat b_{a\oplus b,k}(n-1)$ for the $k$th bit of the $m$th symbol of $\hat \x_{r}(n)$. These received symbols should be demodulated first and then detected by the differential DMF protocol. 

Similar to ML detection used in direct DMF, the differential DMF symbols received at the destination are firstly detected using \eqref{eq:xsml}. After the ML detection, we have the estimated symbol $\hat \x_r(n)$. The equation of this step is similar to \eqref{eq:xsml} and is neglected here. 

The second step is to decode the original data bits of the source symbols $x_a(n-1)$ and $x_b(n-1)$ through an XOR operation as follows
\begin{equation} \label{eq:exdiff}
\begin{aligned}
\hat b_{a,k}(n) = b_{b,k}(n-1) \oplus \hat b_{r, k} (n),\; k=1,2, \\
\hat b_{b,k}(n) = b_{a,k}(n-1) \oplus \hat b_{r, k} (n),\; k=1,2,
\end{aligned}
\end{equation}
where $\hat b_{r, k} (n)$ is the $k$th data bit of the estimated symbol $\hat x_{r} (n)$. The other parameters of the equations are $b_{a,k}(n-1)$ and $b_{b,k}(n-1)$, which are the source data bits transmitted by the corresponding nodes, respectively, at one time slot previously. Such data should be stored in source nodes for later use. We can then obtain the desired data through the XOR operation in \eqref{eq:exdiff}. 

Comparing the error areas of the two DMF schemes, we can find that differential DMF and direct DMF form a complementary relationship: one excels when the other has poor performance. It is easy to know which one should be used for the extreme cases discussed before. However, for a practical fading TWRC model whose channels are fading, the scenario is usually between these extreme situations. Therefore, it is important to find the decision criterion for the appropriate application of HDMF.

% === Optimal decision rule === %

\subsection{Implement HDMF at Relay} \label{sec_opd}

{\color{black}
As discussed above, direct and differential DMFs are more suitable for different channel settings of the TWRC model. A key problem is to apply the  ML decision criterion to choose the right DMF schemes. The Log-Likelihood Ratio (LLR) tool can be used to implement such criterion. 

The LLR value of each bit of the symbols under direct DMF is calculated based on channel conditions and SNR of the received signals, as follows,
\begin{equation}\label{LLR_DDk}
\begin{aligned}
\mathcal{L}_{DirA}(n,k)=\log \frac{P(y_r(n)|b_{a,k}(n)=1)}{P(y_r(n)|b_{a,k}(n)=0)}, \\
\mathcal{L}_{DirB}(n,k)=\log \frac{P(y_r(n)|b_{b,k}(n)=1)}{P(y_r(n)|b_{b,k}(n)=0)},
\end{aligned} 
\end{equation}
where $\mathcal{L}_{DirA}(n,k)$ is the LLR of the $k$th bit of the $m$th symbol from A at the current time slot and $\mathcal{L}_{DirB}(n,k)$ is that obtained from B, where the index $m$ is omitted in the equations for simplicity. 
Similarly, the LLR value under differential DMF can be obtained as follows,
\begin{equation}\label{LLR_DFDk}
\mathcal{L}_{Dif}(n,k)= \log \frac{P(y_r(n)|b_{a,k}(n) \neq b_{b,k}(n) )}{P(y_r(n)|b_{a,k}(n) = b_{b,k}(n) )},
\end{equation}
where $\mathcal{L}_{Dif}(n,k)$ is the LLR of the $k$th bit under the protocol of differential DMF given $x_a(n)$ and $x_b(n)$. 

It is easy to know that the sign of an LLR value under the ML criterion denotes the detected bit result and its absolute value denotes the degree of confidence \cite{Sklar2001DCF}. For a packet containing several symbols, we can try to find the symbol with the minimum confidence, which would most likely cause high Packet Error Rate (PER).

The LLR values of all symbols (where each symbol has $K$ bits) in the received packet at Relay, e.g., $y_r(n)$, ($M$ total symbols)  under direct DMF are calculated as follows,
\begin{equation}\label{LLR_DD}
\begin{aligned}
\mathcal{L}_{DirA}(n)=\sum_{k=1}^K |\mathcal{L}_{DirA}(n,k) |, \\
\mathcal{L}_{DirB}(n)=\sum_{k=1}^K |\mathcal{L}_{DirB}(n,k) |. 
\end{aligned}
\end{equation}
Similarly, the LLR values of differential DMF are calculated as
\begin{equation}\label{LLR_DFD}
\mathcal{L}_{Dif}(n)=\sum_{k=1}^K |\mathcal{L}_{Dif}(n,k) |.
\end{equation}

A typical digital communication system usually uses packets to transmit data and PER can be used as the performance index to denote communication quality. Thus the decision rule for a packet to be processed through direct or differential DMF is given as follows 
\begin{equation}\label{Packet_LLR}
\begin{aligned}
{\rm ~min} & \{|\mathcal{L}^m_{Dir}(n)|,  m =1,...,M \} >  {\rm min}\{|\mathcal{L}^m_{Dif}(n)|, \\
& m =1,...,M \},\;\;\; {\rm Direct ~DMF}, \\
{\rm ~min} & \{|\mathcal{L}^m_{Dir}(n)|,  m =1,...,M \} \leq {\rm min}\{|\mathcal{L}^m_{Dif}(n)|, \\
& m =1,...,M \}, \;\;\; {\rm Differential ~DMF}, \\
& |\mathcal{L}^m_{Dir}(n)| = {\rm max} \{|\mathcal{L}^m_{DirA}(n)|, |\mathcal{L}^m_{DirB}(n)|\}.
\end{aligned}
\end{equation}
We select the smallest LLR value from one packet transmitted from the stronger channel under direct DMF and compare it with the smallest LLR under differential DMF. By using this criterion, the DMF scheme with a greater minimum value will be selected to forward data. Based on PER, even if there is only one failing symbol, the whole packet is labelled as in error. The criterion above therefore ensures that the one with lower PER is selected. 
}

\subsection{HDMF Detection at End Nodes}

The Relay with HDMF protocol can generate and forward two different kinds of messages by using the direct or differential DMF. For the end nodes (A and B), it is essential to be able to detect which kind of messages are transmitted from Relay. A possible solution is to add one bit in the packet header at the Relay, which indicates the DMF scheme used. However, such method increases implementation complexity and changes the frame structure, which is less desirable for practical wireless systems. In this section, we propose a blind detection algorithm with low complexity.

At the end nodes, e.g., source B, the received packet $\y_b(n)$ is first demodulated and mapped to data bits. A Cyclic Redundancy Check (CRC) is then performed on these data bits. If it is correct, this message is processed by direct DMF at Relay. If it is not correct, the second detection attempt using differential DMF (See Section \ref{secDET}) will be carried out. The decoded data bits, e.g., $\hat{b}_{a,k}(n)$, are checked by the CRC. If it is correct, differential DMF is used in the relay. Otherwise, the packet has some error symbols and should be discarded. 

Finally, we analyze the above detection algorithm to see whether it can function properly in real life conditions. For a packet with a large enough number of symbols, e.g., M = 128, it contains 256 bits under QPSK modulation. (Higher order modulation has even more data bits.) {\color{black} We assume the data bits are randomly generated with equal probabilities for 0 and 1. A bit level XOR operation with another randomly generated data packet would change approximately $50\%$ of the data bits. Therefore, for a packet with 256 bits, approximately 128 bits would be changed after the XOR operation. This would identify the difference between direct DMF and differential DMF schemes and the probability of making a mistake is quite low. (If we consider a simple system without source coding and channel coding, the error rate is approximately $1/2^{128}$. With coding, the error rate can be more than this value because of coding dependency.)}

% === Higher order solution ===
\subsection{A Protocol Independent from Modulation Schemes} \label{secHigh}

The hybrid DMF scheme is introduced using QPSK as an example. In order to apply HDMF in a system with any modulation scheme, the following technique is proposed: At the Relay, instead of the usual demodulation schemes, we introduce bitwise detection using the LLR values calculated in \eqref{LLR_DDk} and \eqref{LLR_DFDk}. Detection of the $k$th data bit of the $m$th 
symbol is given below for direct DMF,
\begin{equation}\label{DirDec}
 \hat b_{r,k}(n) =  \left\{ \begin{array}{l}
1, \;{\rm If \;}   \mathcal{L}_{Dir}(n,k) \geq 0; \\
0, \;{\rm If \;}   \mathcal{L}_{Dir}(n,k) < 0.
\end{array} \right.
\end{equation}
If differential DMF is selected through the criterion introduced in Section \ref{sec_opd}, the following detection is used
\begin{equation}\label{DifDec}
\hat b_{r,k} (n)= \left\{ \begin{array}{l}
 1, \;{\rm If \;}   \mathcal{L}_{Dif}(n,k) \geq 0; \\
 0, \;{\rm If \;}   \mathcal{L}_{Dif}(n,k) < 0.
\end{array} \right.
\end{equation}

The Relay will then modulate these detected data bits into symbols following the designated modulation schemes and forward the results to the end nodes. The detection of these symbols at A and B is similar to the case of QPSK (e.g. using the ML algorithm) and the details are neglected here.

\section{Mathematical Analysis} \label{secMA}

This section analyzes the proposed scheme and try to establish the mathematical expression for the end-to-end instantaneous Symbol Error Rate (SER) and obtain the ratios for packets sending through direct and differential DMF. It is easy to know that errors usually come from two phases: the multiple access phase and the broadcast phase. In order to simplify the analysis, we take BPSK as an example. The error rate at relay is denoted by $P_{r}$ and the error rates at the destination node A and B are denoted by $P_{ra}$ and $P_{rb}$ respectively. The end-to-end SER from A to B can be expressed as follows
\begin{equation}
\begin{aligned}
P_{ab} &= 1 - (1-P_r)(1 - P_{rb}) - P_r P_{rb} \\
\end{aligned}
\end{equation}
Similarly $P_{ba}$ can be obtained. The instantaneous SER can be expressed as
\begin{equation}
\begin{aligned}\label{phdmf}
P_{HDMF} &= \frac{1}{2} \left( P_{ab} + P_{ba} \right) \\
	&= P_r + (0.5 - P_r) (P_{ra} + P_{rb}) 
\end{aligned}
\end{equation}

{\color{black}

The average SER of HDMF can be obtained as 
\begin{equation}\label{EPHDMF}
\begin{aligned}
\mE\{P_{HDMF}\} = &\int_{\alpha} \int_{\beta} \int_{\gamma} \int_{\zeta} 
				  P_{HDMF} \\
					&	f(\alpha) f(\beta) f(\gamma) f(\zeta)\;
					\md \alpha \; \md \beta\; \md \gamma\; \md \zeta
\end{aligned}
\end{equation}
where $\alpha = |h_{AR}|$, $\beta = |h_{BR}|$, $\gamma = |h_{RA}|$ and $\zeta = |h_{RB}|$, which follow Rayleigh distributions with the probability density function as $f(x) = \frac{x}{\delta^2} e^{- x^2/(2\delta^2)}$, where $\delta^2$ is the Rayleigh distribution parameter. We can assume these channel variables are independent and identically distributed (i.i.d.), so that the above equation is then simplified to 
\begin{equation}
\begin{aligned}
\mE\{P_{HDMF}\} = \mE\{P_r\} + (0.5 - \mE\{P_r\} ) (\mE\{P_{ra}\} + \mE\{ P_{rb}\}  ), 
\end{aligned}
\end{equation}
where $\mE\{P_{ra}\}$, $\mE\{P_{rb}\}$ and  $\mE\{P_r\}$ are the average SER of the Relay-A channel, Relay-B channel and sources-relay channels, respectively.

%Following we assume all the symbols have the same power, thus the Eb/N0 is the same  

It is easy to obtain the average error rate for the relay to source channels as follows,
\begin{equation}
\begin{aligned}
\mE\{P_{ra}\} &= \int_\gamma P_{ra} f(\gamma)  \md \gamma \\
			  &= \int_0 ^{\infty} Q\left(\sqrt{ \frac{\gamma^2 \epsilon_{ra}} {\sigma_{ra}^2} } \right) f(\gamma) \md \gamma \\
			  & = \frac{1}{2} \left(1 - \sqrt{ \frac{\rho_{ra}}{\rho_{ra} + 1}}\right) 
\end{aligned}
\end{equation}
where $\rho_{ra} = {\epsilon_{ra}}/{\sigma_{ra}^2} $ denotes the signal to noise ratio (SNR) and $Q\{x\} = \frac {1}{\sqrt{2 \pi}} \int _x ^\infty e^{-t^2/2} {\rm d} t $. Similarly
\begin{equation}
\begin{aligned}
\mE\{P_{rb}\} &= \frac{1}{2} \left(1 - \sqrt{ \frac{\rho_{rb}}{\rho_{rb} + 1}}\right). 
\end{aligned}
\end{equation}

The average SER of the sources-relay channels composes of two components from the direct demodulation (A-Relay and B-Relay) and differential demodulation (A$\&$B-Relay),
\begin{equation}\label{PR-1}
\begin{aligned}
\mE\{P_{r}\} = \int_\alpha \int_\beta \left( P_{abr} p_{abr}+P_{ar} p_{ar} +  P_{br} p_{br}   \right) f(\alpha) f(\beta)  \md \alpha \md \beta,
\end{aligned}
\end{equation}
where $P_{ar}$ and $P_{br}$ are the instantaneous SERs of the A-Relay and B-Relay channel, respectively. $p_{ar}$ and $p_{br}$ are the probabilities to choose direct DMF. $P_{abr}$ is the instantaneous SER of the differential DMF and $p_{abr}$ is the probability to choose such scheme. 

For simplicity, \eqref{PR-1} is divided into two parts: the differential DMF part and the direct DMF part. 
\begin{equation}\label{EPr}
\begin{aligned}
\mE\{P_{r}\} = \mE\{P_{Dif}\}  + \mE\{P_{Dir}\},
\end{aligned}
\end{equation}
where
\begin{equation}
\begin{aligned}
\mE\{P_{Dif}\}  &=  \int_\alpha \int_\beta  P_{abr} p_{abr}  f(\alpha) f(\beta)  \md \alpha \md \beta, \\	
	\mE\{P_{Dir}\} &= \int_\alpha \int_\beta \left( P_{ar} p_{ar} +  P_{br} p_{br} \right) f(\alpha) f(\beta)  \md \alpha \md \beta.
\end{aligned}
\end{equation}

Following the same ML criterion used in detecting symbols, the probability to select differential DMF ($p_{abr}$) or direct DMF ( $p_{ar}$, $p_{br}$) can be obtained. Firstly, we consider the ML detection of a differential symbol,
\begin{equation}
\hat x_{a \oplus b}(n)  = {\rm arg} \max_{x_{a \oplus b}(n)  \in \mathcal{M} } \left\{ P(y_r(n)|x_{a\oplus b}(n) \right\},
\end{equation}
the error rate of differential detection is given as 
\begin{equation}
\begin{aligned}
P_{abr} = \frac{1}{2} ( P(\hat x_r^m = 1 | x^m_a \oplus x^m_b = 0 ) +  P(\hat x^m_r = 0 | x^m_a \oplus x^m_b = 1 ))
\end{aligned}
\end{equation}
The above equation can be bounded by [\cite{EPA2010JU}, eq. (7)] as follows,
\begin{equation}\label{DMFbounds}
\begin{aligned}
Q & \left( \sqrt{ 2 \min \left\{ \alpha^2 \rho_a, \beta^2 \rho_b \right\} } \right) <P_{abr} \\
& < Q \left( \sqrt{ {2 \alpha^2 \rho_a }} \right)  + Q\left(\sqrt{{2\beta^2 \rho_b}} \right)
\end{aligned}
\end{equation}

The same rule of $\eqref{Packet_LLR}$ is used to analyze the average SER as follows, 
\begin{equation}\label{pABRAR}
\begin{aligned}
p_{abr} &\approx p\{\phi_{abr} \geq \max \{ \phi_a, \phi_{b} \} \} \\
p_{ar}  &\approx p\{\phi_{a} > \phi_{abr} \cap \phi_a \geq \phi_b \} \\
p_{br}  &\approx p\{\phi_{b} > \phi_{abr} \cap \phi_b \geq \phi_a \} 
\end{aligned}
\end{equation} 
Since the ML criterion is used to obtain \eqref{DMFbounds}, we can derive the parameters following similar rules. Given the exponentially decreasing characteristic of $Q(x)$ function, the two bounds of \eqref{DMFbounds} become tight when SNRs are large. The lower bound of $P_{abr}$ was selected in the calculation for its function to highlight the low-bound of the proposed protocol's SER, 
\begin{equation}
P_{abr} \approx Q  \left( \sqrt{ 2 \min \left\{ \alpha^2 \rho_a, \beta^2 \rho_b \right\} } \right),
\end{equation}
from which we can derive the receiving SNR of an equivalent channel as $\phi_{abr}= 2 \min\{\alpha^2 \rho_a, \beta^2\rho_b \}$. The receiving SNR of the A-Relay channel is $\phi_a = \alpha^2 \rho_a$ for a channel gain of $\alpha^2$, and that of the B-Relay channel is $ \phi_b = \beta^2 \rho_b$.

%Denote SNR of the A-Relay channel as $\phi_{ar}$, SNR of the B-Relay channel as $\phi_{br}$, and SNR of the A$\&$B-Relay equivalent channel as , the probabilities of selecting differential and direct DMF can be obtained as follows

%Similarly the ML rules of the A-R and B-R channel are re-written as
%\begin{equation}
%\begin{aligned}
%\hat x^m_a(n)  &= {\rm arg} \min_{x^m_a(n)\in \mathcal{M}} |y^m_r(n) - h_{AR}x^m_a(n)| \\
%\hat x^m_b(n)  &= {\rm arg} \min_{x^m_b(n)\in \mathcal{M}} |y^m_r(n) - h_{BR}x^m_b(n)|,
%\end{aligned}
%\end{equation}
%from which we can obtain the instantaneous receiving SNR of the A-R channel as $\phi_a = \alpha^2 \rho_a$ for a given channel gain $\alpha^2$ and $ \phi_b = \beta^2 \rho_b$ for the B-R channel, where for the simplicity of analysis, the interference from the other channel is treated as noise in the direct DMF scheme. 
Therefore the following approximation can be obtained,
\begin{equation}\label{selectionprobability}
\begin{aligned}
p_{abr} &\approx p\{\beta^2 \rho_b \geq  \alpha^2 \rho_a \geq  \frac{1}{2} \beta^2\rho_b  \cup \alpha^2\rho_a  \geq  \beta^2 \rho_b \geq  \frac{1}{2} \alpha^2\rho_a \} \\
p_{ar}  &\approx p\{\alpha^2 \rho_a \geq  2 \beta^2\rho_b \}  \\
p_{br}  &\approx p\{\beta^2 \rho_b \geq  2 \alpha^2\rho_a \}  
\end{aligned}
\end{equation}
They are used for the calculation of average SERs.  

%Only the A-R channel has been written here because the B-R channel will be similar and its calculation is not required as demonstrated later. 

It is easy to see that the two parts of $p_{abr}$ following a symmetrical pattern. Since all the channels are i.i.d. and the transmitting power is the same in all nodes, we can apply the symmetry feature to simplify the calculation as follows
\begin{equation}
\begin{aligned}
\mE\{P_{Dif}\}  \approx&  \int_0 ^\infty  \int_{\beta \sqrt{\frac{\rho_b}{2\rho_a}}} ^{\beta \sqrt{\frac{\rho_b}{\rho_a}}} Q\left( \sqrt{2 \rho_a \alpha^2} \right) f(\alpha) f(\beta) \md \alpha \md \beta  \\
& +  \int_0 ^\infty  \int_{\alpha \sqrt{\frac{\rho_a}{2\rho_b}}} ^{\alpha\sqrt{\frac{\rho_a}{\rho_b}}} Q\left( \sqrt{2 \rho_b \beta^2} \right) f(\beta) f(\alpha) \md \beta \md \alpha   	\\
 =& 2 \int_0 ^\infty  \int_{\beta \sqrt{\frac{\rho_b}{2\rho_a}}} ^{\beta\sqrt{\frac{\rho_b}{\rho_a}}} Q\left( \sqrt{2 \rho_a \alpha^2} \right) f(\alpha) f(\beta) \md \alpha \md \beta 
\end{aligned}
\end{equation}

By introducing the full expression of Q function, we have
\begin{equation}
\begin{aligned}
\mE\{P_{Dif}\} \approx& 2 \int_0 ^\infty  \int_{\beta \sqrt{\frac{\rho_b}{2\rho_a}}} ^{\beta \sqrt{\frac{\rho_b}{\rho_a}}} \left[ \frac{1}{\sqrt{2\pi}}  \int_{\sqrt{2 \rho_a} \alpha} ^\infty  e^{-\frac{t^2}{2}} \md t\right] \\
&\times f(\alpha) f(\beta) \md \alpha \md \beta 
\end{aligned}
\end{equation}
Reordering the two integrals of $t$ and $\alpha$, with the new regions of integrals for ${\beta \sqrt{\frac{\rho_b}{2\rho_a}}} \leq \alpha \leq {\beta\sqrt{\frac{\rho_b}{\rho_a}}}$ and $\beta \sqrt{\rho_b} \leq t < \infty$, we have
\begin{equation}
\begin{aligned}
\mE\{P_{Dif}\} \approx& 2 \int_0 ^\infty  \int_{\beta \sqrt{\rho_b}} ^\infty \frac{1}{\sqrt{2\pi}} e^{-\frac{t^2}{2}} \left[\int_{\beta \sqrt{\frac{\rho_b}{2\rho_a}}} ^{\beta\sqrt{\frac{\rho_b}{\rho_a}}} f(\alpha) \md \alpha \right] f(\beta) \md t \md \beta \\
	=& 2 \int_0 ^\infty  \int_{\beta \sqrt{\rho_b}} ^\infty \frac{1}{\sqrt{2\pi}} e^{-\frac{t^2}{2}} \left[e^{-\frac{\rho_b \beta^2}{4 \delta^2 \rho_a} } - e^{-\frac{\rho_b \beta^2}{2 \delta^2 \rho_a} }   \right] f(\beta) \md t \md \beta \\
	=& \mE\{P'_{Dif}\} - \mE\{P''_{Dif}\}. 
\end{aligned}
\end{equation}

The first term of the above equation can be calculated as 
\begin{equation}\label{Pprimedif}
\begin{aligned}
\mE\{P'_{Dif}\} =&  2 \int_0 ^\infty  \int_{\beta \sqrt{\rho_b}} ^\infty \frac{1}{\sqrt{2\pi}} e^{-\frac{t^2}{2}} \frac{\beta}{\delta^2} e^{-\frac{\beta^2}{2 \delta^2} \left( \frac{\rho_b}{2 \rho_a} + 1 \right)  } \md t \md \beta \\
	=&  2 \int_0 ^\infty \frac{1}{\sqrt{2\pi}} e^{-\frac{t^2}{2}} \left[ \int_0 ^\frac{t}{\sqrt{\rho_b}} \frac{\beta}{\delta^2} e^{-\frac{\beta^2}{2 \delta^2} \left( \frac{\rho_b}{2 \rho_a} + 1 \right)  }\md \beta \right]
	   \md  t \\
	=& \frac{4\rho_a}{2\rho_a + \rho_b}  \int_0 ^\infty \frac{1}{\sqrt{2\pi}} e^{-\frac{t^2}{2}} \left(1-e^{-\frac{t^2}{2 \delta^2} (\frac{1}{2\rho_a} + \frac{1}{\rho_b}) }  \right) \md t, \\
	=& \frac{2\rho_a}{2\rho_a + \rho_b}\left( 1 - \frac{1}{\sqrt{1+\frac{1}{2\rho_a \delta^2} + \frac{1}{\rho_b \delta^2}}} \right).
\end{aligned}
\end{equation}
Similarly, the second term can be calculated as
\begin{equation}\label{Pprime2dif}
\begin{aligned}
\mE\{P''_{Dif}\} = \frac{\rho_a}{\rho_a + \rho_b}\left( 1 - \frac{1}{\sqrt{1+\frac{1}{\rho_a \delta^2} + \frac{1}{\rho_b \delta^2}}} \right).
\end{aligned}
\end{equation}

For the direct DMF part, with the same assumptions of i.i.d. fading and the symmetrical regions of $p_{ar}$ and $p_{br}$, the two channels should have the same SER as follows
\begin{equation}
\begin{aligned}
\mE\{P_{Dir}\} &\approx 2 \int_\alpha P_{ar} p_{ar} \md \alpha.
\end{aligned}
\end{equation}
where the instantaneous SER of A-Relay channel under the ML criterion can be easily obtained as $P_{ar} = Q(\sqrt{\frac{\alpha^2}{\beta^2 + 1/\rho_a}})$ \cite{Proakis83}. Following the previous assumption of high SNR, an approximation is obtained as $P_{ar} \approx  Q({\frac{\alpha }{\beta }})$ and then the average SER is given as follows 
\begin{equation}
\begin{aligned}
\mE\{P_{Dir}\} &\approx 2 \int_0^\infty  \int_{\sqrt{\frac{2 \rho_b}{\rho_a} }\beta} ^\infty  Q\left(\frac{\alpha }{\beta }\right)  f(\alpha) f(\beta) \md \alpha \md \beta \\
	&= 2 \int_0^\infty  \int_{\sqrt{\frac{2 \rho_b}{\rho_a} }\beta} ^\infty  \left[ \frac{1}{\sqrt{2\pi}}  \int_{\frac{\alpha}{\beta}} ^\infty  e^{-\frac{t^2}{2}} \md t\right]  f(\alpha) f(\beta) \md \alpha \md \beta \\
\end{aligned}
\end{equation}
The regions of integration for $\alpha$ and $t$ are ${\sqrt{\frac{2 \rho_b}{\rho_a} }\beta} \leq \alpha < \infty$ and $\frac{\alpha}{ \beta} \leq t < \infty$. Reordering these two integrals we have
\begin{equation}\label{pdir}
\begin{aligned}
&\mE\{P_{Dir}\} \approx \\
	& 2 \int_0^\infty \int_{\sqrt{\frac{2 \rho_b}{\rho_a}}} ^{\infty} \frac{1}{\sqrt{2 \pi}} e^{-\frac{t^2}{2}} \left[ \int_{\beta\sqrt{\frac{2 \rho_b}{\rho_a}}}^{\beta t} \frac{\alpha}{ \delta^2} e^{-\frac{\alpha^2}{2 \delta^2}} \md \alpha \right] f(\beta)  \md t \md \beta \\
	& = 2 \int_0^\infty \int_{\sqrt{\frac{2 \rho_b}{\rho_a}}} ^{\infty} \frac{1}{\sqrt{2 \pi}} e^{-\frac{t^2}{2}} \left[ e^{-\frac{\rho_b \beta^2}{\delta^2 \rho_a}}-e^{- \frac{\beta^2 t^2}{2 \delta^2}}   \right] f(\beta) \md t \md \beta, \\	
	&= \mE\{P'_{Dir}\} + \mE\{P''_{Dir}\}
\end{aligned}
\end{equation}
where the regions of $\alpha$ and $t$ are transformed to $\sqrt{\frac{2 \rho_b}{\rho_a}} < t < \infty$ and ${\beta\sqrt{\frac{2 \rho_b}{\rho_a}}} \leq \alpha \leq \beta t$. 

Reordering the two integrals within the above equation, the first term of \eqref{pdir} can be written as
\begin{equation}\label{Pprimedir}
\begin{aligned}
\mE\{P'_{Dir}\} &  \\
 = & 2 \int_{\sqrt{\frac{2 \rho_b}{\rho_a}}} ^{\infty}  \frac{1}{\sqrt{2 \pi}} e^{-\frac{t^2}{2}} \left[ \int_0^\infty \frac{\beta}{\delta^2} e^{-\frac{\beta^2}{2\delta^2 } \left(\frac{2 \rho_b}{\rho_a} + 1\right)  }   \md \beta \right] \md t \\	
= & 2 \int_{\sqrt{\frac{2 \rho_b}{\rho_a}}} ^{\infty}  \frac{1}{\sqrt{2 \pi}} e^{-\frac{t^2}{2}} \frac{1}{\frac{2 \rho_b}{ \rho_a} + 1} \md t \\
= & \frac{2 \rho_a}{2 \rho_b + \rho_a} Q \left( \sqrt{\frac{2 \rho_b}{\rho_a}} \right)
\end{aligned} 
\end{equation}
where we have used the change of variable in the first step.

The second term of \eqref{pdir} can be similarly calculated by reordering the two integrals of $\beta$ and $t$,
\begin{equation}
\begin{aligned}
\mE\{P''_{Dir}\} &  \\
 = & -2 \int_{\sqrt{\frac{2 \rho_b}{\rho_a}}} ^{\infty}  \frac{1}{\sqrt{2 \pi}} e^{-\frac{t^2}{2}} \left[ \int_0^\infty \frac{\beta}{\delta^2} e^{-\frac{\beta^2}{2 \delta^2} (1+t^2)} \md \beta \right] \md t  \\
 = & -2 \int_{\sqrt{\frac{2 \rho_b}{\rho_a}}} ^{\infty}  \frac{1}{\sqrt{2 \pi}} e^{-\frac{t^2}{2}} \frac{1}{t^2+1} \md t  \\
\end{aligned} 
\end{equation}
In order to obtain a closed form of the integration, lower edge of the integration region is applied to obtain the following bound,
\begin{equation}\label{Pprime2dir}
\begin{aligned}
\mE\{P''_{Dir}\} \geq & \frac{-2}{1 + \frac{2 \rho_b}{ \rho_a}} Q \left( \sqrt{\frac{2 \rho_b}{\rho_a}} \right).
\end{aligned} 
\end{equation}
It is easy to see that the decreasing rate of the exponential function $e^{-\frac{t^2}{2}}$  is much faster than $\frac{1}{t^2 + 1}$ with the increase of $t$. The value above thus provides a close approximation.  

With the availability of \eqref{Pprimedif}, \eqref{Pprime2dif}, \eqref{Pprimedir} and \eqref{Pprime2dir}, the average SER of the source to relay channel \eqref{EPr} can be directly obtained. The results are given in Fig.\ref{TheorySimuSER}. 
}

{ \color{black}

The closed forms for the average value of the selection probabilities \eqref{selectionprobability} and its preceding equation \eqref{pABRAR} provides a general measure about the ratio of symbols through differential DMF and direct DMF. They can reveal the underlying properties of the protocol. From the analysis in \eqref{selectionprobability} and the PDFs of $\alpha$ and $\beta$, the average value of $p_{abr}$ can be obtained as
\begin{equation}
\begin{aligned}
\mE\{p_{abr}\} &=  \int_0^{\infty} \int_{\beta \sqrt{\frac{\rho_b}{2\rho_a}}} ^ {\beta \sqrt{\frac{\rho_b}{\rho_a}}}  f(\alpha) f(\beta) \md \alpha \md \beta 
+  \int_0^{\infty} \int_{\alpha \sqrt{\frac{\rho_a}{2\rho_b}}} ^ {\alpha \sqrt{\frac{\rho_a}{\rho_b}}}   f(\beta) f(\alpha) \md \beta \md \alpha \\
&= \frac{2\rho_a}{2\rho_a + \rho_b}- \frac{\rho_a}{\rho_a + \rho_b} + \frac{2\rho_b}{2\rho_b + \rho_a}- \frac{\rho_b}{\rho_a + \rho_b}. 
\end{aligned}
\end{equation}
Similarly, we can obtain the closed forms for the average value of A-Relay channel and B-Relay channel: 
\begin{equation}
\begin{aligned}
\mE\{p_{ar}\} &= \int_0^{\infty} \int_{\beta \sqrt{\frac{2\rho_b}{\rho_a}}} ^ {\infty}  f(\alpha) f(\beta) \md \alpha \md \beta  = \frac{\rho_a}{\rho_a + 2 \rho_b},
\end{aligned}
\end{equation}
\begin{equation}
\begin{aligned}
\mE\{p_{br}\} &= \int_0^{\infty} \int_{\alpha \sqrt{\frac{2\rho_a}{\rho_b}}} ^ {\infty}   f(\beta) f(\alpha) \md \beta \md \alpha = \frac{\rho_b}{2 \rho_a + \rho_b}.
\end{aligned}
\end{equation}

}

\section{Transmission Scheduling over Fading TWRCs and Queue Length Analysis}\label{secTSF}

The two wireless channels of TWRC normally suffer from random fading which deteriorates the communication quality and reduces channel capacity if it is not well handled. Due to the random feature of fading, different channels of TWRC observe different instant fading, and it is therefore usually the case that they have different strengths. Under such conditions, the traditional scheduling schemes, e.g., the scheduling scheme proposed for DNC \cite{Chen13DNC}, which assume equal channel gains, incur large queues at Relay and decrease bandwidth efficiency. On the other hand, with the introduction of HDMF in the TWRC model, a novel scheduling scheme should be proposed to effectively use its potential.

{\color{black}

We analyze the average queue lengths at the end nodes and relay and their instability. Firstly, define the following transmission modes for the whole system:
\begin{itemize}
\item Mode I: A and B transmit data packets to the relay simultaneously at the data rate of $\mathcal{R}_1$ and $\mathcal{R}_2$ using the HDMF protocol;
\item Mode II: Relay broadcasts packets to A and B at the data rate of $\mathcal{R}_3$ if both of its queues have data;
\item Mode III: Relay only transmits packets to B using the data rate $\mathcal{R}_4$;
\item Mode IV: Relay only transmits packets to A using the data rate $\mathcal{R}_5$;
\end{itemize}
}

A scheduling scheme is then proposed where the relay is set to be the administrative node. The scheduling details are given below:  
\begin{enumerate}
\item The Relay chooses a transmission mode from the above four modes with probability $f_i, (i= \{1,2,3,4\})$.
\item If Mode I is selected, both A and B will be instructed by Relay to transmit at the beginning of the time slot. If either of their buffers are not empty, e.g., $\min\{Q_{a}(t), Q_{b}(t)\}>0$, the node with data in its buffer will transmit packets to the relay during this time slot; otherwise, they remain silent. 
\item If Mode II is selected, Relay will broadcast to A and B if both of their queues are not empty i.e., $Q_{ra}$ and $Q_{rb}$ are greater than 0. In the case of one empty queue, Relay will broadcast the data from this queue. If both of them are empty, Relay will remain silent. 
\item The Relay will transmit to B if Mode III is selected and $Q_{rb}$ is not empty. If Mode IV is selected, Relay will transmit to A if $Q_{ra}$ is not empty. If the queue for data to be sent to A or B at Relay is empty when the corresponding mode is selected, Relay will remain silent during this time slot.  
\end{enumerate}

\subsection{Queue Length Analysis}
{\color{black}
Without loss of generality, we consider the route from A to B via Relay. The backward route from B to A through Relay is similar to the forward one and will be omitted here. As described before, the queue length of A is denoted by $Q_{a}$ and the queue length of Relay for the packets to be transmitted to B is denoted by $Q_{rb}$. The state transition relation of $\{Q_{a}, Q_{rb}\}$ can be modelled by a Markov chain using one-step state transition probability $q_{ \{m,k \}, \{i,j\} }$ which denotes the probability of the event that the state changes from $Q_{a}(t) = m $, $Q_{rb} (t) =k$ to $Q_{a}(t+1) = i $, $Q_{rb}(t+1) =j$. The probability that $i$ packets arrive at one source node during the current time slot can be modelled by a Poisson distribution as follows
$$a_i = \frac{(\lambda T)^i}{ i! } \exp(-\lambda T),$$ where $\lambda$ is the data arrival frequency and $T$ is the length of one time slot. An analysis method similar to \cite{Chen13DNC} is introduced here. 

Under the condition of fading TWRC, channel quality can affect the data rate and queue length. The data rate which can be supported by the channel from A to Relay is denoted as $n$ and the probability to support such rate is given by $c_n(t)$. Similarly, the probability for Relay-A channel supporting a data rate of $n$ is denoted as $r_n(t)$, and the probability for Relay-B channel supporting $n$ is $q_n(t)$. %They can be calculated by using Rayleigh distribution. 

%Assuming that channel fading follows a Rayleigh distribution, $c_n(t)$, $r_n(t)$ and $q_n(t)$ can be calculated by applying the required date rate to the Rayleigh distribution function.   

Under Mode I, with the application of HDMF protocol, Relay has two possible inputs: packets from differential DMF or direct DMF. Based on the analysis in Section \ref{secMA}, we can use $p_{abr}$ to denote the probability with which the packets arriving from A (and B) by differential DMF, $p_{ar}$ to denote the probability of packets arriving from A by direct DMF and $p_{br}$ to denote the probability of packets arriving from B by direct DMF. It is easy to know that $p_{abr} + p_{abr} + p_{br} = 1$. The state transition probability is then given by
\begin{equation}\label{qModeI}
\begin{aligned}
&q^{I}_{\{m,k \}, \{i,j\}} =  \\
 &\left\{ \begin{array}{ll}
a_i 		&	m=0, j=k \\
a_i \sum_{n=m}^{\infty}c_n  (p_{abr} + p_{ar}) 	& 	0<m\leq n, j= k+m \\
a_{i-m}  \sum_{n=m}^{\infty}c_n p_{br}	&	0<m\leq n, j=k \\
a_{i-m+n} c_n (p_{abr}+p_{ar}) & m > n \geq 0, j=k+n \\
a_{i-m} c_n p_{br} &	m > n \geq 0, j=k
\end{array}
\right.
\end{aligned}
\end{equation}

Notice that Mode I is selected with the probability of $f_1$ at the beginning of every time slot. The first item in \eqref{qModeI} denotes the state transition probability when the queue at A is empty: $Q_{a}(t)=m=0$. Therefore $Q_{rb}(t+1) = Q_{rb}(t)$ (which means $j=k$). The corresponding probability is $a_i$. 

However, if the queue at A is not empty, the state transition at Relay will depend on the HDMF protocol. If the channel can support a data rate $n\geq m$, the probability of receiving and storing $m$ packets in the queue $Q_{rb}(t+1)$ (and thus causing the corresponding state transition) is $a_i \sum_{n=m}^{\infty}c_n (p_{abr} + p_{ar})$, because both differential DMF and direct DMF can generate the $m$ packets. As a result, the queue length of Relay becomes $j=k+m$. The probability for the relay state remaining $k$ is $a_{i-m}  \sum_{n=m}^{\infty}c_n p_{br}$, because Relay only receives data from B and these data would contribute to none of the two queues under concern. 

On the other hand, if $n < m$, only $n$ packets will be received by Relay through either direct or differential DMF. Therefore, the probability of Relay's state becoming $k+n$ is $a_i \sum_{n=m}^{\infty}c_n (p_{abr} + p_{ar})$. Another case is direct DMF for data from B, which would not contribute to the state transition. So $j=k$ with a probability of $a_{i-m} c_n p_{br}$. 

Under Mode II - Relay broadcasts data to A and B, the state transition probability is given as follows
\begin{equation}\label{qModeII}
q^{II}_{\{m,k \}, \{i,j\}} = \left\{ \begin{array}{ll}
a_{i-m} \sum_{n=k}^{\infty}r_n 	&	j=0 \\
a_{i-m} r_n 					&	j=k-n, 0 \leq n < k
\end{array}
\right.
\end{equation}
If the channel can support a data rate of $n$ which is greater than the number of packets within Relay's buffer $k$, the whole buffer will be transmitted to the destination and thus emptied, e.g. $j=0$. The probability for such state transition is $a_{i-m} \sum_{n=k}^{\infty}r_n $. Otherwise, the new queue length of Relay will become $k-n$, and the state transition probability is $a_{i-m}r_n$.
  
Under Mode III, Relay transmits packets to B and the state transition probability is as follows,
\begin{equation}\label{qModeIII}
q^{III}_{\{m,k \}, \{i,j\}} = \left\{ \begin{array}{ll}
a_{i-m} \sum_{n=k}^{\infty}q_n 	&	j=0, n\geq k\\
a_{i-m} q_n 					&	j=k-n, 0 \leq n < k
\end{array}
\right.
\end{equation}
If the data rate $n$ equals or is greater than the data in the queue at Relay, $j$ will be zero after transmission since all of its data has been transmitted. Otherwise, the queue length will be $k-n$. The corresponding transition probabilities are given in the equation above. 

If Mode IV is selected, Relay will only transmit data to A. The state of our concerned queues will remain the same. Thus the state transition probability is as follows
\begin{equation}\label{qModeIV}
q^{IV}_{\{m,k \}, \{i,j\}} = a_{i-m}, \;\;\; j=k.
\end{equation}

Notice that Relay selects the current transmission mode from the four candidates based on the probability of $f_i\; (i=1,...,4)$. By combining the one-step state transition probabilities and the mode choice probabilities, we can obtain the final transition probability, summarized in  \eqref{qModeTotal}. \addtocounter{equation}{1} The stationary state of the queue lengths of interest can be obtained given the one-step transition probabilities.  

\newcounter{tempequationcounter}
\begin{figure*}[!t]
\normalsize
\setcounter{tempequationcounter}{\value{equation}}
\begin{IEEEeqnarray}{rCl}
\setcounter{equation}{52}
q_{\{m,k \}, \{i,j\}} = \left\{ \begin{array}{ll}
a_i 										&	j=k=m=0 \\
a_{i-m}(1 - f_1 + f_1 c_0)  				& 	j=k=0, m>0 \\
a_{i-m}(f_2 r_n + f_3 q_n)   				&	j=k-n, k>n\geq 0 \\
a_{i-m} (f_2 \sum_{n=m}^{\infty} r_n + f_3 \sum_{n=m}^{\infty} q_n) & j=0, k>0 \\
a_i (f_1+f_2r_0 + f_3 q_0 + f_4)  			& 	j=k>0, m=0 \\
a_{i-m} (f_1p_{br} + f_2 r_0 + f_3 q_0 + f_4) 	&	j=k>0, m>0 \\
a_{i-m+n} c_n (p_{abr}+p_{ar}) f_1						&	j=k+n, m>n \geq 0 \\
a_i \sum_{n=m}^{\infty} c_n (p_{abr}+p_{ar}) f_1			&	j=k+m, m>0 \\
0											& 	else
\end{array}
\right.
\label{qModeTotal}
\end{IEEEeqnarray}
\setcounter{equation}{\value{tempequationcounter}}
\hrulefill
\vspace*{4pt}
\end{figure*}
}

In order to know the stationary-state queue length of $Q_{a}$ at A and $Q_{rb}$ at Relay, we need to work out the stationary-state distribution of the previous mentioned Markov chain with the transition probabilities summarized in \eqref{qModeTotal}. Firstly we rewrite the one-step transition probabilities using the matrix form. By denoting the buffer size for the queue at Relay for B as a big number $N$, the $(x,y)$th element of the transition matrix, e.g. $(\P)_{x,y}$, is given by $q_{\{mN+k, iN+j \}}$ which is equivalent to $q_{\{m,k \}, \{i,j\}}$. The stationary-state distribution of the queue lengths can be obtained as follows
\begin{equation}
\mathbf{\Pi} = \mathbf{1} \mathbf{\cdot} (\I- \P + \U)^{-1},
\end{equation}
where $\mathbf{1}$ denotes a row vector with elements of ones, $\U$ is an all-unity matrix and $\I$ is the identity matrix \cite[15.107]{KobayashiETAL2012}. $\mathbf{\Pi}$ denotes the stationary probability of the two queue lengths $\{ Q_{a}(t), Q_{rb}(t)\}$ and can be used to calculate the average queue length of A, as follows
\begin{equation}
Q_{a} = \sum_{i=0}^{\infty} \sum_{j=0}^{\infty} i \pi _{in+j},
\end{equation} 
and the average queue length of Relay for B, as follows  
\begin{equation}
Q_{rb} = \sum_{i=0}^{\infty} \sum_{j=0}^{\infty} j \pi _{in+j}.
\end{equation}

%=== Simulation === %

\section{Simulation}\label{secSim}

\subsection{Experimental Conditions}

We investigate the performance of the proposed HDMF protocol and the transmission scheduling scheme in this section. {\color{black} The channels are modelled as block Rayleigh distributions and the noise is modelled as additive white Gaussian distributions for most of the cases, except the second experiment in Section \ref{secPer} where compare the protocols under Gaussian channel settings.} We assume reciprocal channel settings, e.g. $h_{AR} = h_{RA}$ and $h_{BR} = h_{RB}$. The modulation scheme used in the nodes is QPSK  and each packet has 128 symbols. At every experiment, a total of $10^5$ packets are transmitted from one source node to the other via Relay.

\subsection{Performance under Relative Channel Strengths}\label{secPer}

\begin{figure}[tb]\centering
\centering
\epsfig{file=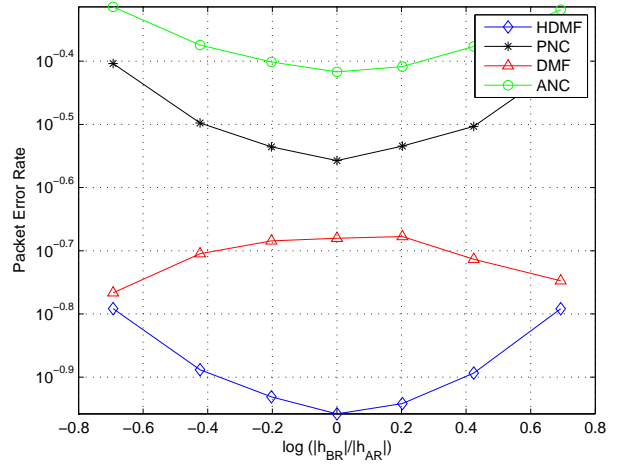,width=1\columnwidth}
\caption{PER performance comparison of the protocols under Rayleigh channel settings and $E_b/N_0 = 15dB$. }
\label{figPerRayleighHdmfPncDmf15dB}
\end{figure}

\begin{figure}[tb]\centering
\centering
\epsfig{file=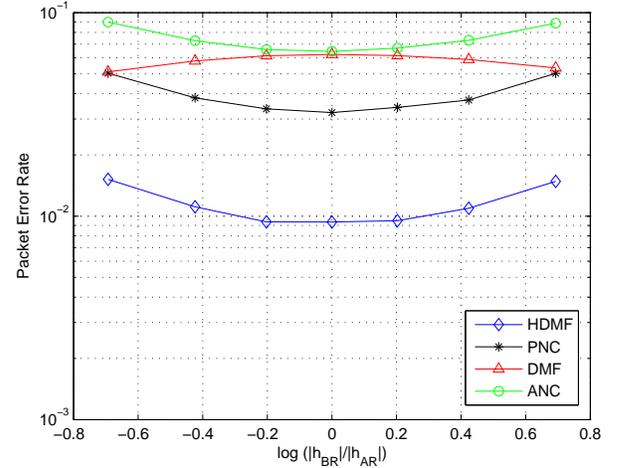,width=1\columnwidth}
\caption{PER performance comparison of the protocols under Rayleigh channel settings and $E_b/N_0 = 25dB$. }
\label{figPerRayleighHdmfPncDmf25dB}
\end{figure}

The first experiment tests the Packet Error Rate (PER) performance of the HDMF protocol, the PNC protocol, the classic DMF protocol and the ANC protocol in Rayleigh fading scenarios. The average channel gains are adjusted using $\log ( \mE\{|h_{BR}|\}/ \mE\{ |h_{AR}|) \}$.  We set the range of this value from  -0.7 to 0.7 and 0 means $\mE\{|h_{BR}|\} = \mE \{|h_{AR}|\}$. Fig.\ref{figPerRayleighHdmfPncDmf15dB} shows that ANC and PNC have the highest PERs, which means their performance is heavily affected by the channel realizations. They also have their own minimum PER when both the two channels have the same average channel gains, e.g.,  $\mE\{|h_{BR}|\} = \mE \{|h_{AR}|\}$. Classical DMF is able to achieve lower PER than ANC/PNC in this case. The proposed HDMF protocol follows a similar tread as ANC/PNC but with lower PER than all of them. Such trend becomes even clearer at high SNR settings, as shown in Fig.\ref{figPerRayleighHdmfPncDmf25dB}, where the SNR is 25dB. In this scenario, the performance of all four protocols is improved; the most impressive change is from PNC as its PER is below DMF. HDMF has the best performance among them. 

\begin{figure}[tb]\centering
\centering
\epsfig{file=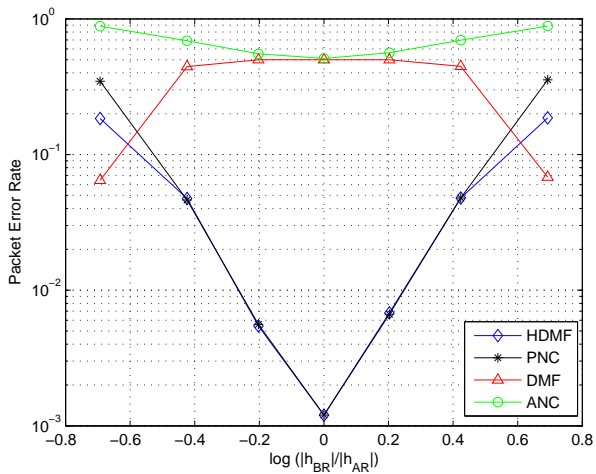,width=1\columnwidth}
\caption{PER performance comparison of the protocols under Gaussian channel settings, $E_b/N_0 = 10$dB. }
\label{figPerGaussianHRAHRB}
\end{figure}

\begin{figure}[tb]\centering
\centering
\epsfig{file=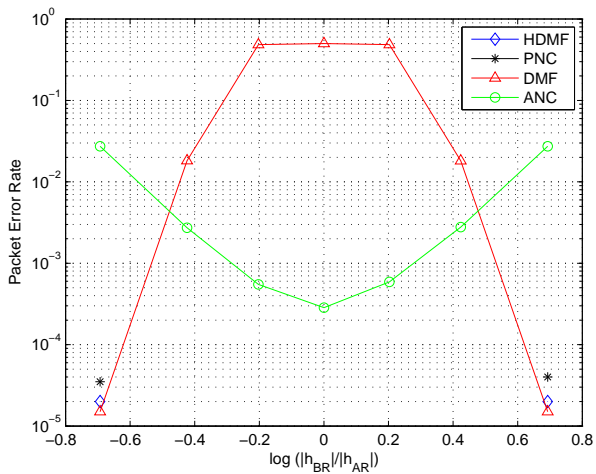,width=1\columnwidth}
\caption{PER performance comparison of the protocols under Gaussian channel settings, $E_b/N_0 = 15$dB. }
\label{figPerGaussianHRAHRB15dB}
\end{figure}

The second experiment investigates the performance of these protocols under the Gaussian channel settings where the fading is small and merges into noise. { \color{black} Since the fading effect is avoided in this experiment, the impact of noise towards these protocols can be clearly identified.} The channel gains vary by following the relative strength of $\log (|h_{BR}|/ |h_{AR}|)$. A similar range is set from -0.7 to 0.7, where 0 means $|h_{BR}| =|h_{AR}|$. Figs. \ref{figPerGaussianHRAHRB}, \ref{figPerGaussianHRAHRB15dB}, \ref{figPerRayleighHdmfPncDmf15dB} and \ref{figPerRayleighHdmfPncDmf25dB} show the experimental results in terms of PER.

From Fig. \ref{figPerGaussianHRAHRB}, we can see that under Gaussian channel settings and SNR per bit $E_b/N_0=$  $10$dB, HDMF and PNC achieve similar performance at the middle of the relative channel strength range ($~[-0.4,0.4]$), with its lowest PER at the point where the two channels have the same strength. Such advantages decrease when the channel gains of the two channels (A-Relay and B-Relay) diverge and finally they are overtaken by the DMF protocol. As expected, DMF has poor performance when the two channels have similar strengths because one channel heavily interferes the demodulation of the other. ANC has the worst performance because it amplifies noise significantly, especially when SNR is low. We raise SNR from 10dB to 15dB and obtain Fig. \ref{figPerGaussianHRAHRB15dB}. Under this new setting, the performance of ANC is improved remarkably and it has lower PER than DMF. The PERs of HDMF and PNC are also improved significantly and have similarly low value at most of the x-axis range {\color{black} (They are 0 and not shown in the figure)}.  

{\color{black}

\subsection{Average SER}

This section compares the SER performances of the different relaying protocols. The experiment conditions are similar to the previous experiments, except for the following settings: 1) the average fading channel gains are fixed to 1; 2) $E_b/N_0$ varies from 5dB to 30dB; 3) both the theoretical analysis and simulation use BPSK modulation. 

Fig.\ref{SERalll5T30dB} demonstrates the average SERs of the four protocols. All of them have decreasing SER when SNR increases. However, at the lower SNR region ($E_b/N_0<10$dB), DMF has lower SER than the other protocols. With the increase of SNR, the SERs of HDMF and PNC drop much faster than DMF. After $25$dB, even ANC has a lower SER than DMF. For the regions with SNR $>10$dB, HDMF outperforms PNC, ANC and DMF.  

\begin{figure}[tb]\centering
\epsfig{file=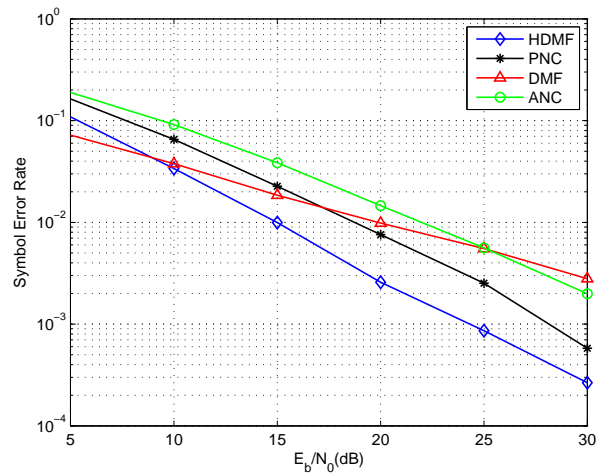,width=1\columnwidth}
\caption{SER performance of the four protocols. }
\label{SERalll5T30dB}
\end{figure}

Fig.\ref{TheorySimuSER} compares the average SER performance between simulation and theoretical analysis \eqref{EPHDMF}. From the figure, we can see that simulation result matches well with theoretical analysis, particularly at high SNR regions for the reason of applying high SNR assumption in the theoretical analysis. Such results confirm the performance of the proposed protocol.  
\begin{figure}[tb]\centering
\epsfig{file=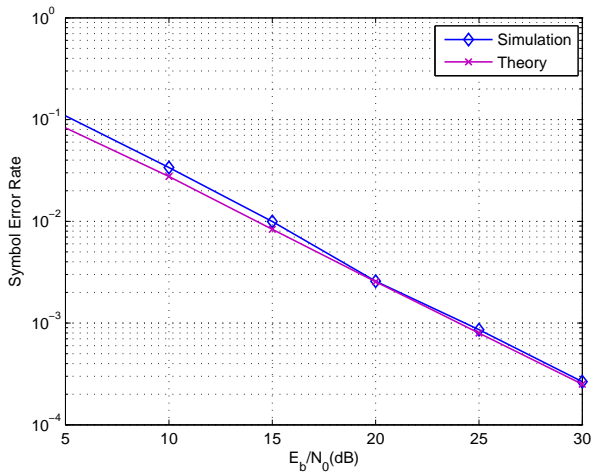,width=1\columnwidth}
\caption{Average SER comparison. }
\label{TheorySimuSER}
\end{figure}

}

\subsection{Queue Length Analysis}

This experiment studies the queue length of A and the queue length for data to B at Relay in terms of $Q_{a}$ and $Q_{rb}$. We assume the mode selection frequency $f_{i}$ to be equal for all the four modes, e.g., $f(i)=1/4, i=1,...,4$. The channels are modelled as independent Rayleigh fading channels, e.g., $c_n$, $r_n$ and $q_n$ follow the Rayleigh probability distribution function. Similar to \cite{Chen13DNC}, the packet arrival frequency for A is $\lambda = 0.5$ frames/slot. The time slot spans 1ms, e.g., $T=1$ms, and the SNR is $20$dB. We introduce the same design logic as in \cite{Chen13DNC} to configure the system to accommodate a higher average arrival rate than the actual rate of $\lambda: \rightarrow \lambda(1+ \epsilon)$ for the purpose of reducing memory usage and reasonable running time, where $\epsilon$ has a small positive value. 

\begin{figure}[tb]\centering
\epsfig{file=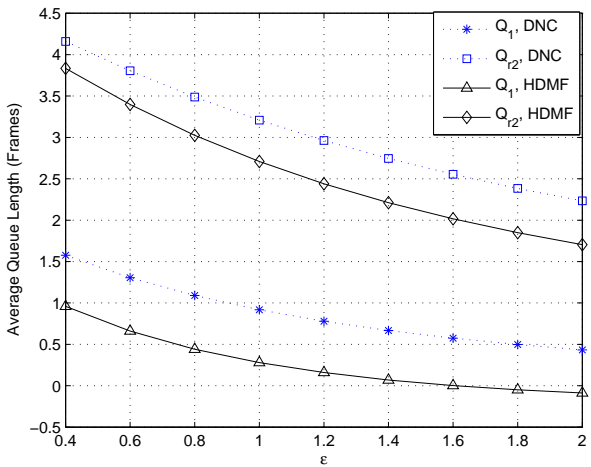,width=1\columnwidth}
\caption{Stationary-state queue length of A and Relay for B versus $\epsilon$. Rayleigh fading channels are used. The two protocols HDMF and DNC are compared at the packet arriving rate of 0.5 frame/slot.}
\label{figQueueLength}
\end{figure} 

Fig. \ref{figQueueLength} shows the stationary-state queue lengths for the proposed HDMF and the referenced scheduling for Digital Network Coding (DNC) \cite{Chen13DNC}. It can be seen from the figure that the proposed HDMF has comparable smaller queue lengths both at Relay and node A than that of DNC. The reason is that HDMF utilizes the two advantages: reduced time slots consumption and opportunistic DMF under different channel conditions. LLR controls the protocol switching probabilities (i.e., $p_{abr}$, $p_{ar}$ and $p_{br}$) to ensure the packet detection and effective packet forwarding. Generally, the queue length of HDMF is about 0.5 frames less than the DNC protocol. With the increase of $\epsilon$, the queue lengths of both protocols reduce significantly. As $\epsilon$ is introduced to adjust the system throughput and capacity, however, the decrease in queue length occurs with the cost of decreased system throughput since there will be more idle slots during which no data is sent out.

% === Discussion ===

\section{Discussion}\label{secDis}

%\subsection{Multi-hop networks}

This section discusses some issues involving the application of the HDMF protocol. The first issue is about multi-hop networks with several intermediate nodes. The demodulation and implementation of HDMF should be similar to the system model in this paper. However, data processing on relays should be carefully handled in order to avoid the same packet being sent back to its origin, thus reducing the efficiency. The {\color{black} routing} algorithms developed for multihop sensor networks, e.g., \cite{Gharavi03MUT}, can be applied to solve this problem. (The specific implementation details are omitted in this paper.)

%\subsection{Packets with different lengths}

This paper introduces the HDMF protocol with the default setting that the packets from two source nodes have the same length. If they are of different lengths, HDMF can still be applied with small amendments. Because the detection at Relay is based on LLR values, we can add some {\color{black} extra 0 bits} to the shorter packet in order to calculate the LLR values. A better enhancement is to use these bits for the purpose of error-correcting coding\cite{Proakis83}.

%\subsection{Synchronization}

ANC and PNC can perform well if the two channels of TWRC have similar gains. This places a high requirement on synchronization as practical wireless channels cannot always guarantee the simultaneous arrival of signals from two distributed sources, thus the synchronization of Relay to one source would mean discrimination against the other and cause significant detection errors. If such case does happen, the proposed HDMF would be reduced to direct DMF automatically and avoid the negative influence from the severely discriminated channel which would degrades PNC or ANC significantly, and is thus less vulnerable to synchronization errors. 

% implementation of the protocol
The proposed HDMF does not try to solve the fundamental problem of the signal to interference ratio being too low, rather it avoids the problem by relaying the signals through differential DMF rather than detecting the information forcefully which might be degraded by interference and noise. In this case, differential DMF provides more information for destination to be able to complete the final symbol detection, e.g. destination can reduce its own contribution in the interference. 

{ \color{black}
% end-to-end BER
The instantaneous end-to-end error rate can provide a more accurate measure for the relay to make decision. However, its cost includes overheads incurred by feedbacks from the nodes and increased computation complexity. The next stage of the research will investigate these factors and try to establish an efficient solution based on the current protocol framework and the instantaneous end-to-end error rate. 
}

%\subsection{Limits of HDMF}

%The most significant disadvantage of HDMF is the limited data rate of the feedback route. Extensive experiments show that approximately $60\%$ of packets comparing to forward link can be sent to source by the HDMF protocol. As a result, it is not suitable for scenarios with equal bidirectional traffic.  

%As shown in Section \ref{secRat}, 

% Next step
%  \subsection{Theorectical proofs}

% === Conclusion  === %

\section{Conclusion}\label{secCon}

%Mobile and wireless networks have become the primary or even sole access method for more and more people, it is expected to have increased examples of TWRC models where multiple users exchange data through intermediate nodes. Furthermore, rapid increase of traffics from mobile users makes effective and efficient network protocols critical for both the operators and users in terms of performance and revenue. 

This paper proposes an HDMF protocol for the TWRC model where existing protocols are troubled by high error rate or difficulty in synchronization. We study the components of HDMF and the fundamentals of direct DMF, differential DMF and the key detection criterion. We further analyze the queue length of the model with HDMF protocol. Comparison of its performance with the existing protocols and scheduling scheme indicates that the proposed HDMF has a lower average PER and smaller average queue length.

% as part of future work, we plan to ...

\section*{Acknowledgement}\label{ack}
This paper is sponsored by the Research Council UK Digital Economy Theme Sustainable Society Network+ and Royal Society-NSFC Grant No. IE131036. The authors would like to thank the anonymous reviewers for their helpful comments and suggestions, particularly on the hybrid-DF and HDMF comparison, average SER and selection probabilities analysis, queue analysis and synchronization. 

%This work is funded by the project

\bibliographystyle{IEEEtran}
\bibliography{lcb}

\end{document}